\documentclass[aps,prb,superscriptaddress,showpacs,reprint,floatfix]{revtex4-2}
\usepackage{physics}
\usepackage{amsmath,amssymb}
\DeclareMathOperator{\sgn}{sgn}
\usepackage{bbm}
\usepackage{amssymb}
\usepackage{makeidx}
\usepackage{amsfonts}
\usepackage{subfigure}
\usepackage{epstopdf}
\usepackage{pst-grad} 
\usepackage{mathtools}
\usepackage[colorlinks,hyperindex]{hyperref}
\usepackage[inline]{trackchanges}

\newcommand{\HF}{\mathcal{H}_\text{F}}

\hypersetup
{	colorlinks,%
	citecolor=green,%
	linkcolor=red,%
	urlcolor=blue%
}
\allowdisplaybreaks[4]

\newcommand{\lettersection}[1]{\emph{#1}.---}
\newcommand{\Hfull}{H_\text{full}}

\begin{document}


\title{Engineering subharmonic responses beyond prethermalization via Floquet scar states}

\author{Ke Huang}
\affiliation{Department of Physics, City University of Hong Kong, Kowloon, Hong Kong SAR, China}
\author{Xiao Li}
\email{xiao.li@cityu.edu.hk}
\affiliation{Department of Physics, City University of Hong Kong, Kowloon, Hong Kong SAR, China}

\begin{abstract}
In this work we propose a new scheme to engineer subharmonic responses via scar states in a generalized PXP model. 
We first show that the generalized PXP model also possesses a band of scar states like the pristine PXP model does. 
In addition, we reveal that a generalized forward scattering approximation (FSA) still works for these scar states. 
We further argue that the FSA subspace exhibits an $SO(3)$ symmetry, which enables an $SO(3)$-FSA approach for the scar states. 
When such a model is placed under periodic driving, nontrivial Floquet scar states emerge in the quasienergy spectrum. 
One appealing feature of such Floquet scar states is that they can support subharmonic responses akin to the discrete time crystalline phase. 
In particular, such subharmonic responses can exist beyond the conventional prethermalization regime, where either large frequencies or large driving amplitudes are required. 
\end{abstract}

\maketitle


\lettersection{Introduction}
Ergodicity in an isolated quantum system is believed to be described by the eigenstate thermalization hypothesis (ETH)~\cite{Deutsch1991,Srednicki1994,Rigol2008,Mori_2018}, assuming that eigenstates around the same energy are locally indistinguishable. 
However, abundant theoretical and experimental evidences suggest that some system breaks the ETH. 
A well-known example is many-body localization~\cite{Nandkishore2015,Altman2015,Abanin2019}, which breaks ergodicity in the strong sense: all eigenstates violate the ETH, so all localized initial states retain their memories even after a long period of time evolution.  
Meanwhile, the ETH can also be weakly broken by embedding rare atypical states in the thermal bulk~\cite{Shiraishi2017}, and only dynamics from particular initial states exhibit nonergodic behavior. 
Experimentally, the weak ergodicity breaking was first found in the strongly interacting Rydberg atoms~\cite{Bernien2017}, in which only quench dynamics from the $Z_2$ ordered states revive periodically. 
This phenomenon is theoretically described by a special set of rare eigenstates in the PXP model~\cite{Turner2018_NatPhys,Turner2018}. 
The periodic revival resembles the scar in the chaotic systems, and thus the rare eigenstates in the PXP model are dubbed quantum many-body scar (QMBS) states. 

Beyond isolated systems, people also found nonergodic dynamics in periodically driven or Floquet systems. 
A prominent example is the discrete time crystal (DTC) stabilized by strong disorders~\cite{Khemani2016_PRL,Else2016,Yao2017,Zhang_2017,Choi_2017,Randall2021,Mi2021,Frey2022,Xu2021}, in which all initial states exhibit subharmonic responses with respect to the driving period. 
Meanwhile, there is the so-called prethermal DTC, in which some ergodic initial states may also behave like DTC within an exponentially long time scale~\cite{Abanin2017,Else2017,Maskara2021_PRL,Kyprianidis2021}. 
Such examples are typically explained within the framework of spontaneous discrete symmetry breaking. 
Recently, another possibility of ergodicity breaking in Floquet systems emerges, which are known to have Floquet QMBS states by analogy with the isolated systems. 
A trivial case is that the Hamiltonian at different time shares a set of common QMBS states that spontaneously become the Floquet QMBS states. 
In contrast, several recent works have proposed nontrivial Floquet QMBS states~\cite{Pai2019_PRL,Mukherjee202006_PRB,Mukherjee202008_PRB,Mizuta2020_PRR,Yarloo2020_PRB,Sugiura2021_PRR,Rozon2021_arXiv,Haldar2021,Biao2022,Hudomal2022}. 
However, such proposals are typically limited by specific parameters or require a high-frequency driving. 

In this work, we generalize the PXP model to a three-component model, whose QMBS states can also be approximated by the forward scattering approximation (FSA)~\cite{Turner2018_NatPhys,Turner2018}. 
In fact, the generalized PXP model completes the approximate $su(2)/so(3)$ Lie algebra of the pristine PXP model~\cite{Turner2018,Choi2019_PRL,MondragonShem2021} in the FSA subspace.
As a result, the evolution of the system represents the Lie group generated by this Lie algebra. 
We verify numerically that the Lie group is the $SO(3)$ Lie group instead of the $SU(2)$ Lie group.  
Facilitated by the group structure, the evolution of a time-dependent generalized PXP model emulates a three-dimensional rotation.  
Particularly, one can readily engineer a large variety of Floquet systems with different subharmonic revivals (period doubling, tripling, etc), where the system spontaneously possesses nontrivial Floquet scar states determined by the group structure. 
Despite the similarity to the prethermal discrete time crystal (DTC), we emphasize that this model is fundamentally different, because the generalized PXP model depends on a unbroken continuous symmetry rather than a broken discrete symmetry.

\lettersection{The generalized PXP model}
We start by reviewing the pristine PXP Hamiltonian~\cite{Turner2018_NatPhys,Turner2018}
\begin{align}
    H_x =\sum_r\sigma_r^x \qty(\dfrac{1}{2}+Q_r)\prod_{\langle r,r'\rangle}P_{r'},
\end{align}
where $\sigma_r^x,\sigma_r^y,\sigma_r^z$ are Pauli matrices at site $r$, and $P_r=(\mathbbm{1}+\sigma_r^z)/2$ projects into the excited state. 
In addition, $Q_r$ serves as stabilizer to enhance the weak ergodicity breaking in this model. 
Though different forms of $Q_r$ have been proposed for different lattices~\cite{Choi2019_PRL,Michailidis2020}, the condition $[Q_r,\sigma_{r'}^z]=0$ for all $r,r'$ is always satisfied, which is crucial for the discussion below. 

The QMBS states in the PXP model have an equal energy spacing of $\omega_s=2\pi/T_s$ (where we set $\hbar=1$ throughout), which is almost independent of the system size.
As the pristine PXP Hamiltonian includes only the $x$-component of the spin operators, it is intuitive to introduce the other two components in a 1D bipartite lattice as well, which are given by 
\begin{align}
    H_y&=\sum_r\sgn(r)\sigma_r^y\qty(\dfrac{1}{2}+Q_r)\prod_{\langle r,r'\rangle}P_{r'},\\
    H_z &=\dfrac{\omega_{s}}{2}\sum_r\sgn(r)\sigma_r^z\prod_{\langle r,r'\rangle}P_{r'}.
\end{align}
In the above Hamiltonian, the lattice is divided into two sublattices $A$ and $B$, and we introduce
\begin{align}
    \sgn(r) = 
    \begin{cases}
        1, & r \in A\\
        -1, & r \in B
    \end{cases}. 
\end{align}
Thus, for an arbitrary unit vector $\vec n=(n_x,n_y,n_z) \in \mathbb{R}^3$, we can construct a generalized PXP Hamiltonian as 
\begin{align}
    \Hfull = \vec{n}\cdot\vec{H},
    \qq{where} 
    \vec H&=(H_x,H_y,H_z). 
     \label{Eq:H_Full} 
\end{align}
In what follows, we will study the scar states in this generalized PXP model. 
For convenience, we will also use the notation that 
\begin{align}
\vec n=(\sin\theta\cos\varphi,\sin\theta\sin\varphi,\cos\theta) \equiv (\theta,\varphi). 
\end{align}

\begin{figure}[t]
\includegraphics[width=\columnwidth]{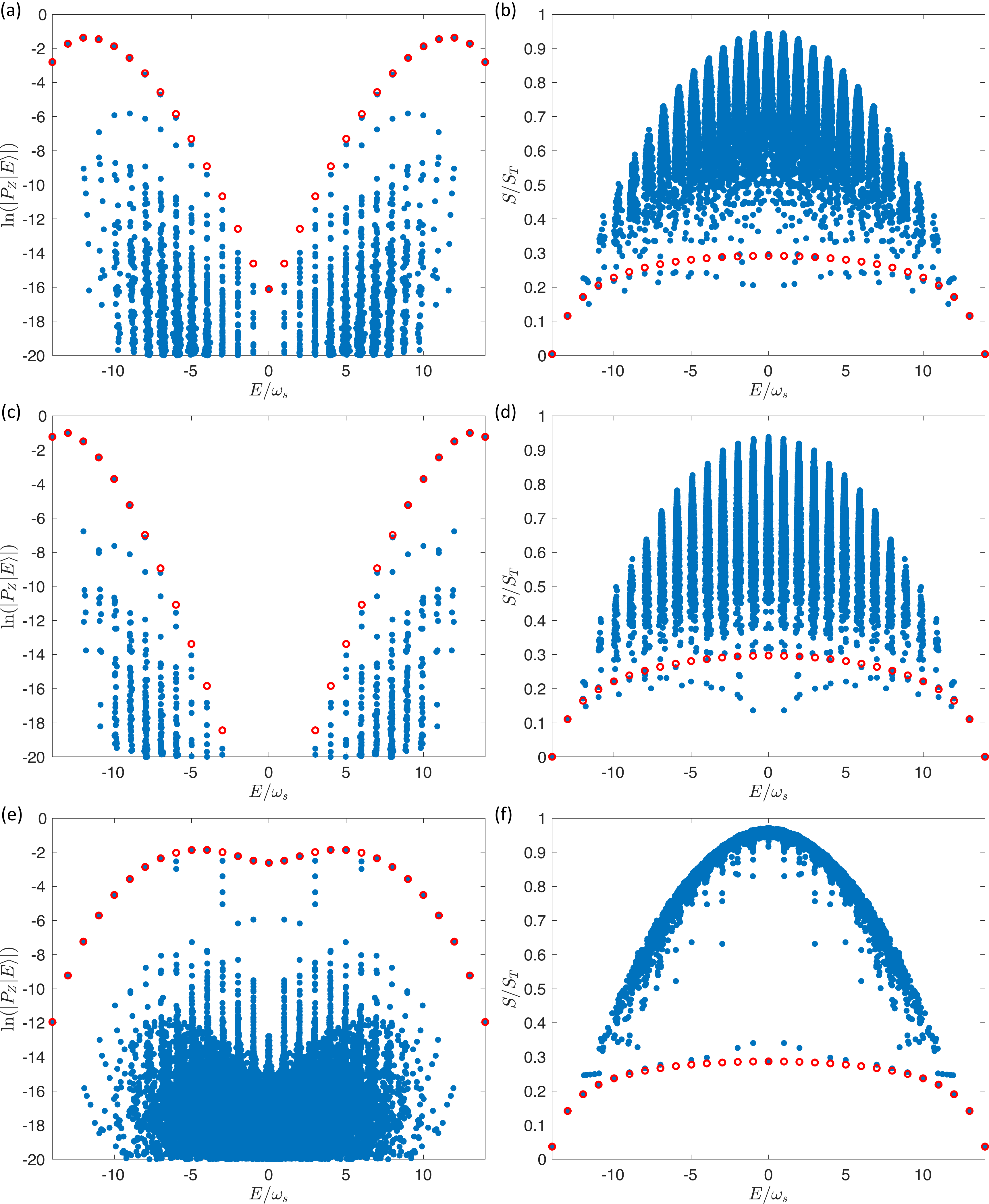}
\caption{\label{Fig:OverlapEE}
Left panel plots the projection, and the right panel plots the half-chain EE in an $L=28$ chain. 
Here we take $\vec n=(\pi/5,0)$ (large $\abs{n_z}$) in (a) and (b), $\vec n=(\pi/5+2\pi/3,0)$ (large $\abs{n_z}$) in (c) and (d), and $\vec n=(2\pi/5,\pi/2)$ (small $\abs{n_z}$) in (e)  and (f). 
The EE in the thermal limit is $S_T=\ln(F_{L/2+2})-1/2$ for 1D chains, where $F_n$ is the $n$th Fibonacci number.
}
\end{figure}

\lettersection{Scar states in the generalized PXP model}
We now demonstrate the existence of scar states in $\Hfull$ in Eq.~\eqref{Eq:H_Full}. 
To begin with, note that a bipartite lattice permits two maximally excited states $\ket{M_\mathrm{A}}$ and $\ket{M_\mathrm{B}}$, defined as the state where  all sites on the $A$ or $B$ sublattice are excited, respectively. 
Let us denote the subspace spanned by these two states as $\mathbb{M}$. 
The scars of the system can be identified by their overwhelmingly large projection on $\mathbb{M}$, and their significantly small entanglement entropy (EE), defined as $S(|\phi\rangle)=-\mathrm{Tr}\left\{\rho_\mathrm{L}(|\phi\rangle)\ln\rho_\mathrm{L}(|\phi\rangle)\right\}$. 
In this definition, the system is divided into two subsystems: the left and right half, and $\rho_\mathrm{L}(|\phi\rangle)=\mathrm{Tr}_\mathrm{R}(|\phi\rangle\langle\phi|)$. 
These properties are shown in Fig.~\ref{Fig:OverlapEE}, which implies the existence of scars in $\Hfull$. 

A hallmark of the pristine PXP model is that the QMBS states can be approximated by the so-called forward-scattering approximation (FSA)~\cite{Turner2018_NatPhys,Turner2018}, which considers a specific subspace generated by a pair of ladder operators,
\begin{align}
    H^+=H_x+iH_y, \quad 
    H^-=H_x-iH_y,
\end{align}
and the basis of this FSA subspace is
\begin{equation}
\begin{aligned}
    \ket{j}=&(H^+)^j\ket{M_\mathrm{A}} \times \norm{(H^+)^j\ket{M_\mathrm{A}}}^{-1},\\
    \ket{N-j}=&(H^+)^j\ket{M_\mathrm{B}}\times \norm{(H^+)^j\ket{M_\mathrm{B}}}^{-1},\\
    \ket{N/2}=&\left(\ket{N/2-1}+\ket{N/2+1}\right)\\&\times\norm{\ket{N/2-1}+\ket{N/2+1}}^{-1},
\end{aligned}
\end{equation}
where $j=0, 1, \cdots, (N/2)-1$, and $N$ is the number of sites. 
This idea can also be applied to the generalized PXP model as follows. 
Consider the FSA Hamiltonian given by $P_F \Hfull P_F$, where $P_F=\sum_{j}\dyad{j}$ projects into the FSA subspace. 
As in the pristine PXP model, the FSA Hamiltonian here is tridiagonal. 
In addition, it has nonzero diagonal entries because of $H_z$. 
In Fig.~\ref{Fig:OverlapEE} we can see clearly that FSA provides an accurate approximation as expected, and the approximation works better for small $\abs{n_z}$. 

\begin{figure}[!]
\includegraphics[width=0.8\columnwidth]{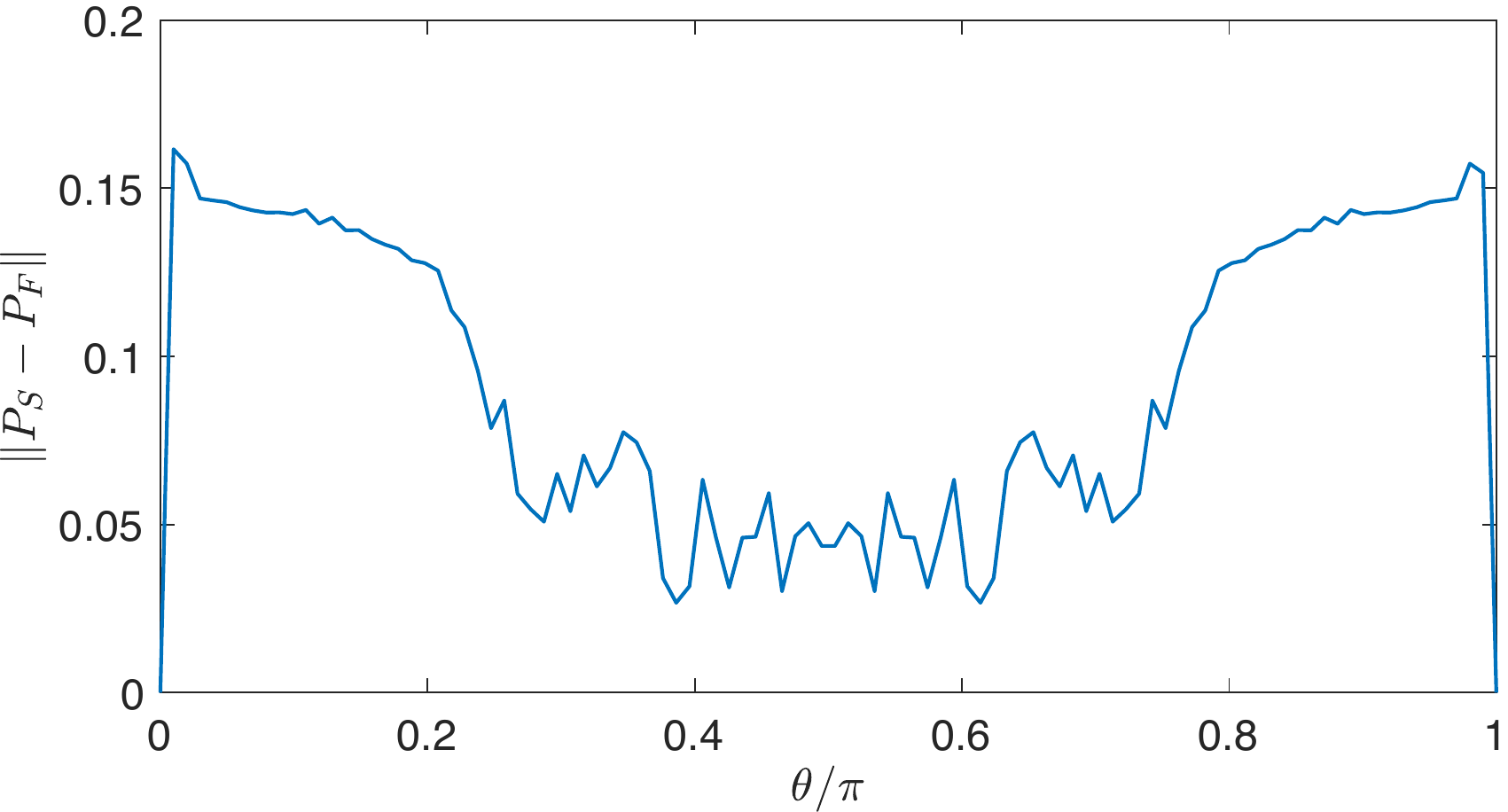}
\caption{\label{Fig:ScarSubspace}
A plot of $\norm{P_S-P_F}$ as a function of $\theta$. 
We thus see that the FSA subspace provides a good approximation for the scar subspace. Here, we take $L=18$.}
\end{figure}

Moreover, this generalized FSA implies that the scar subspaces of different $\vec n$  can be approximated by a $\vec n$-independent subspace, the FSA subspace.
To quantify this assertion, we first denote $\{\ket{E_j(\vec n)}\}$ as the set of scars of $\Hfull$. 
The projector of the scar subspace is then $P_S(\vec n)=\sum_{j}\dyad{E_j(\vec n)}$. 
Further, we introduce the normalized Frobenius norm, 
\begin{align}
    \norm{X}\equiv\mathrm{Tr}\left(X^\dag X\right)/(N+1).
\end{align}
Under this convention, we have $\norm{P_S(\vec n)}=\norm{P_F}=1$, and the difference between the FSA subspace and the scar subspace can be estimated by $\norm{P_S(\vec n)-P_F}$. 
Additionally, $\norm{P_S(\vec n)-P_F}$ is $\phi$-independent because of $[H_z,P_F]=0$ and the exact $z$-axis symmetry 
\begin{equation}
\begin{aligned}
    \qty[R_z(\alpha)\vec n]\cdot\vec H=U^\dag\qty(\vec n\cdot\vec H)U, \label{z_symmetry}
\end{aligned}
\end{equation}
where $U=e^{-i \alpha H_z/\omega_{s}}$ and $R_z(\alpha)$ is the rotation around $z$-axis for an angle $\alpha$. 
As a result, we will only consider $\vec{n}$ with $\varphi=0$ (or $n_y = 0$) henceforth without loss of generality. 
Figure~\ref{Fig:ScarSubspace} shows that the FSA subspace indeed provides a good approximation for the scar subspace, with an overall accuracy $\norm{P_S-P_F}<0.17$. Besides, it also indicates that the FSA subspace is more accurate for small $\abs{n_z}$, consistent with Fig.~\ref{Fig:OverlapEE}.
Therefore, we have demonstrated that the FSA subspace is approximately an invariant subspace for all $\Hfull$. 

More interestingly, $\Hfull$ possesses an approximate $su(2)$ Lie algebra on the FSA subspace~\cite{Turner2018,Choi2019_PRL,MondragonShem2021}, i.e., 
\begin{align}
    \comm*{P_FH_aP_F}{P_FH_bP_F} \approx i\varepsilon_{abc}\,\omega_{s}P_FH_cP_F,
\end{align}
where $a,b,c=x,y,z$. 
Hence, what remains to resolve is which symmetry group ($SO(3)$ or $SU(2)$) can best describe the FSA subspace. 
The first hint is that the dimension of the FSA subspace is always odd because we consider \emph{bipartite lattices} only, implying that it may be a representation of $SO(3)$. 
Furthermore, one can show that for arbitrary unit vectors $\vec n_1,\vec n_2$ and rotation angles $\omega_sT_1,\omega_sT_2$, there exist $\vec n_3$ and $T_3$ satisfying
\begin{align}
    R(\vec n_2,\omega_{s}T_2)R(\vec n_1,\omega_{s}T_1)&=R(\vec n_3,\omega_{s}T_3),\\
    P_Fe^{-iT_2\vec n_2\cdot\vec H}e^{-iT_1\vec n_1\cdot\vec H}P_F&\approx P_Fe^{-iT_3\vec n_3\cdot\vec H}P_F, \label{SO(3)}
\end{align}
where $R(\vec n,\alpha)$ represents the rotation around $\vec n$ for an angle $\alpha$. 
As a result, the approximate symmetry of the FSA subspace is $SO(3)$ instead of $SU(2)$ in this bipartite lattice. 
We will see that this approximate $SO(3)$ symmetry of the FSA subspace is crucial for our design of the subharmonic response when the generalized PXP model is placed under periodic driving.

\lettersection{The generalized PXP model under periodic driving}
In the pristine PXP model, the presence of QMBS states with equal energy spacing results in weak ergodicity breaking. 
This phenomenon can also be observed in the Floquet PXP model, as we now show. 
Specifically, we study the dynamics from the initial state $\ket{\mathbb{Z}_2}=\ket{M_\mathrm{A}}$~\footnote{However, we note that the physics we discuss applies to all initial states in the scar subspace.}, and introduce a time-periodic Hamiltonian $H(t)$ with a period of $T=T_1+T_2$, 
\begin{align}
H(t)=
\begin{cases}
 \vec n_1\cdot\vec H, &  0<t<T_1\\
 \vec n_2\cdot\vec H, &  T_1<t<T_1+T_2
\end{cases}, 
\label{Eq:Floquet_Hamiltonian}
\end{align}
where $\vec n_1, \vec n_2$ are unit vectors in $\mathbb{R}^3$.
As the FSA subspace approximates the true scar subspace for all $\vec n$, the FSA Hamiltonian $P_FH(t)P_F$ is applicable to the dynamics of $\ket{\mathbb{Z}_2}$ in the Floquet system as well. 
In particular, under FSA, the time evolution operator is approximated by 
\begin{align}\label{Eq:FSA}
    U(t) \approx U_\text{FSA}(t) \equiv  
    \mathcal{T}\exp[-i\int_0^tP_FH(s)P_F \,\dd{s}], 
\end{align}
where $\mathcal{T}$ is the time ordering operator.

\begin{figure}[!]
\includegraphics[width=0.8\columnwidth]{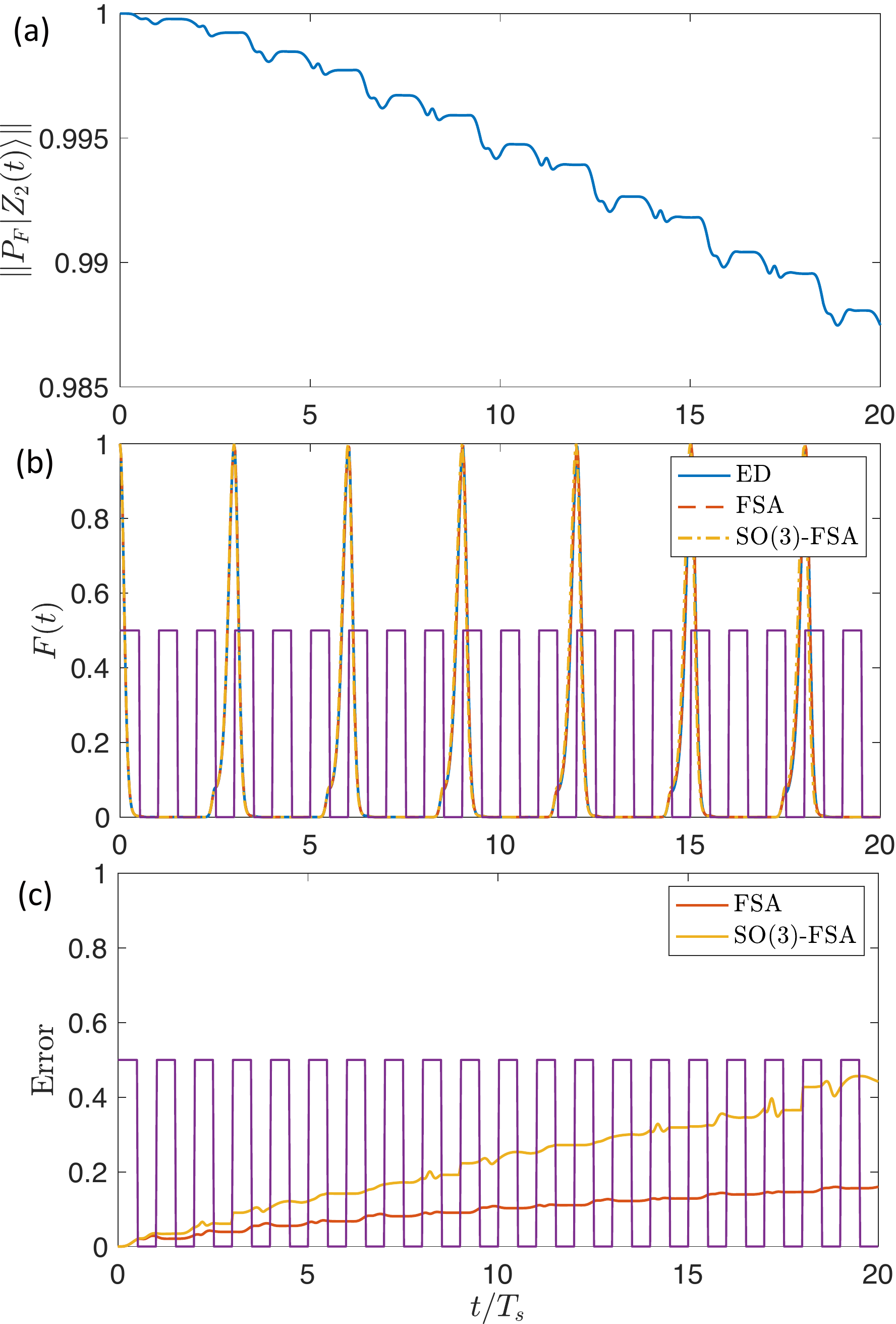}
\caption{\label{Fig:Dynamics_DTC}
(a-c) plot the projection in the FSA subspace, the fidelity, and the accuracy of two FSA's, respectively. Here we use an $L=28$ chain ($T_{s}=9.8987$), with parameters $\vec n_1=(\pi/5,0)$, $\vec n_2=(\pi/5+2\pi/3,0)$, $T_1=T_s/2$, and $T_2=T_s/2$. 
The purple line in (c) equals $1/2$ when $\vec n_1\cdot\vec H$ is turned on, and equals zero otherwise. }
\end{figure}

From the $SO(3)$ symmetry perspective in Eq.~\eqref{SO(3)}, we know that $H(t)$ executes a rotation $R(t)$ in $\mathbb{R}^3$. 
Knowing that $R(t)$ can be uniquely expressed as $R(t)=R(\vec n_3(t),\omega_sT_3(t))$, where $\vec n_3(t)$ and $T_3(t)$ are continuous functions, we introduce an $SO(3)$-FSA approximation for $U(t)$, given by  
\begin{align}\label{Eq:SO(3)FSA}
    U(t) \approx U_\text{$SO(3)$-FSA}(t) \equiv e^{-iT_3(t) \vec n_3(t)\cdot P_F\vec H P_F}.
\end{align}
Given that a revival is equivalent to $R(t)$ becoming the identity, we are able to engineer the period of revival to be \emph{any} integer multiple of the driving period.

For definiteness, here we demonstrate a period tripling case, and leave the period quadrupling case to the Appendix. 
We start by showing that $\ket{\mathbb{Z}_2(t)}$ still almost stays within the FSA subspace even for $t=20T_s$, as shown in Fig.~\ref{Fig:Dynamics_DTC}(a). 
Then, the fidelity $F(t)=\abs{\braket{\mathbb{Z}_2}{\mathbb{Z}_2(t)}}^2$ calculated through FSA and $SO(3)$-FSA is compared with that obtained by the exact diagonalization (ED) method. 
We find that both $SO(3)$-FSA and FSA offer a good approximation after a long time evolution (up to $t \sim 20 T_s$), as shown in Fig.~\ref{Fig:Dynamics_DTC}(b). 
This verifies our assertion that the FSA subspace has an approximate $SO(3)$ symmetry. 

We also study the error of the two approximations by evaluating the norm between the approximated states and the ED results, which are shown in Fig.~\ref{Fig:Dynamics_DTC}(c).
We find that the FSA generally fits the ED results fairly well. 
In contrast, the difference between the $SO(3)$-FSA approximation and the ED results continues to increase.  
We emphasize that the error depends strongly on $t/T_s$, but not very much on the driving period $T$ of the model $H(t)$.

The observed periodic revivals suggest the existence of Floquet QMBS states in the system described by Eq~\eqref{Eq:Floquet_Hamiltonian}. 
In particular, they are the eigenstates of a Floquet Hamiltonian $\HF$, defined by way of 
\begin{align}
e^{-i T \HF} \equiv  U(T) = e^{-iT_2\vec n_2\cdot\vec H}e^{-iT_1\vec n_1\cdot\vec H} . 
\end{align} 
What is more, the Floquet QMBS states cannot be trivial because $\vec n_1\cdot \vec{H}$ and $\vec n_2\cdot \vec{H}$ generally share no common eigenstates. 
Notwithstanding, as the FSA subspace is approximately invariant, Eq.~\eqref{Eq:FSA} suggests that $\HF$ is approximated by the Floquet Hamiltonian $\HF^\text{FSA}$ defined by $e^{-i T \HF^\text{FSA}} \equiv U_{\text{FSA}}(T)$, which is expected to capture the QMBS states. 
Meanwhile, Eq.~\eqref{Eq:SO(3)FSA} implies that under $SO(3)$-FSA, $\HF$ is approximated by 
\begin{align}
&P_F\HF P_F\nonumber\\
\approx& \HF^{\text{$SO(3)$-FSA}}\equiv\dfrac{T_3(T)}{T} P_F\left[\vec n_3(T)\cdot\vec H\right]P_F. \label{SO3-FSA}
\end{align}
The importance of this approximation is to explain the periodic revivals in our model, as not all QMBS states support periodic revivals. 


\begin{figure}[!]
    \includegraphics[width=\columnwidth]{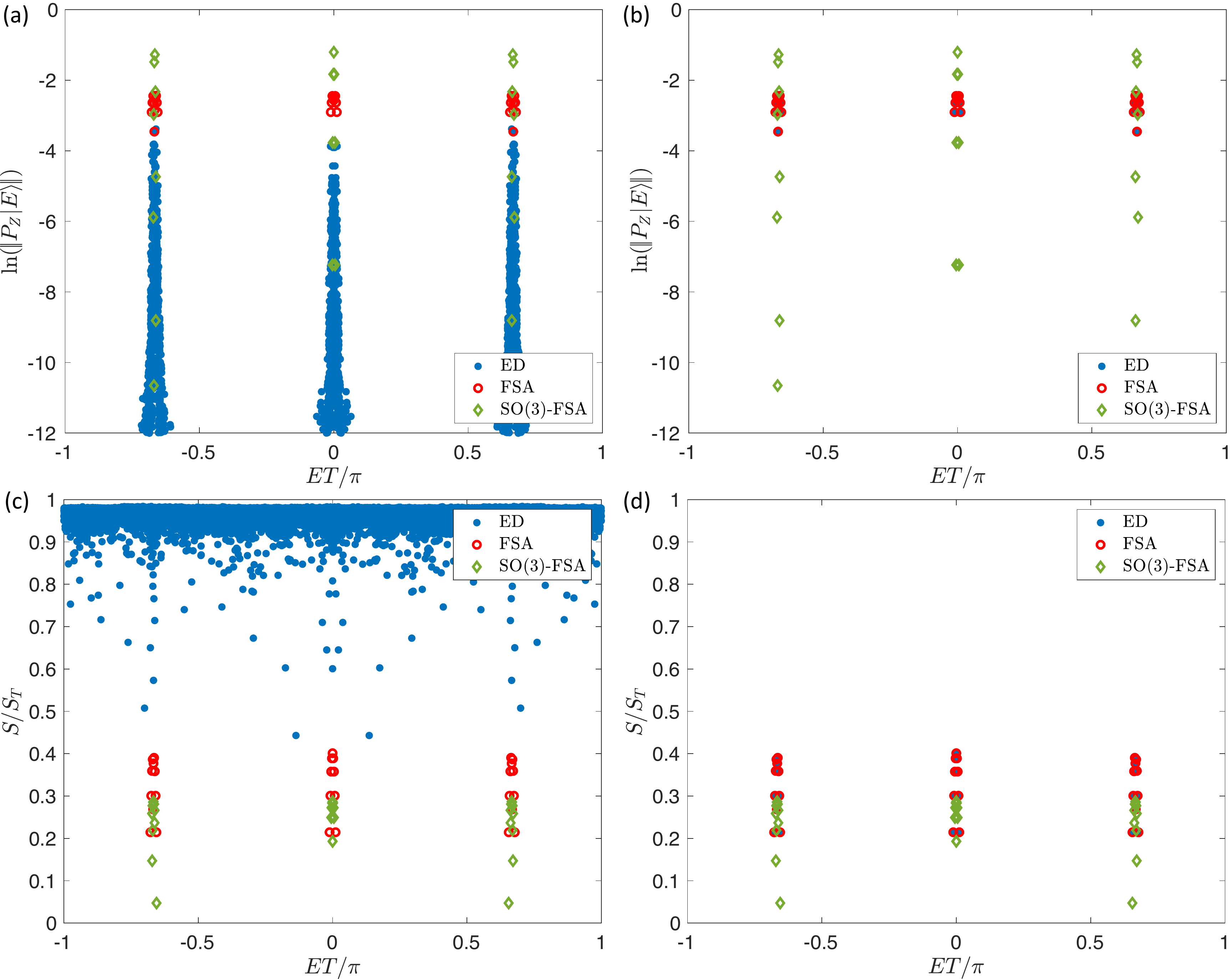}
    \caption{\label{Fig:Fscar_overlap_DTC} The projection and the EE of the eigenstates derived from three different approaches. The ED results in the left and the right panel represent the eigenstates of $U(T)$ and $P_F U(T) P_F$, respectively. The parameters are given in Fig.~\ref{Fig:Dynamics_DTC}.}
\end{figure}

In Fig.~\ref{Fig:Fscar_overlap_DTC}, we study the eigenstates of $U(T)$ and $P_FU(T) P_F$ by ED and compare them with the results obtained by FSA and $SO(3)$-FSA. 
Noting the equivalence between $ET=\pm\pi$ in the quasi-energy spectrum, we find in Fig.~\ref{Fig:Fscar_overlap_DTC}(a) that the eigenstates of $U(T)$ form a tower structure, and that they concentrate within a narrow energy window around the quasienergies given by the FSA and the $SO(3)$-FSA approaches. 
However, neither approximations predicts the projection $\ln\qty(\norm{P_Z\ket{E}})$ very well. 
The reason is that the model does not have actual eigenstates serving as the QMBS states. 
Instead, the system possesses several towers of eigenstates, which can be regarded as bands of degenerate states within a small time scale. 
Therefore, before full thermalization, the superposition of each tower of states plays the role of QMBS states. 
However, we emphasize that because of the extremely small bandwidth, the relaxation time here is still rather long with respect to the typical time scale $T_s$, as shown in Fig.~\ref{Fig:Dynamics_DTC}(a). 
Furthermore, Fig.~\ref{Fig:Dynamics_DTC}(a) suggests that the short-time evolution happens essentially within the FSA subspace. 
Therefore, the superposition should be described by the eigenstates of $P_F U(T) P_F$ calculated in Fig.~\ref{Fig:Fscar_overlap_DTC}(b). 
For the subharmonic responses in our model, we observe that the eigenstates of $P_F U(T) P_F$ can be perfectly approximated by the FSA method, whereas the $SO(3)$-FSA only gives the correct energy but not the projection. 
The reason is that the spectrum in this case is highly degenerate. 
Consequently, the imperfection of the symmetry on the FSA subspace serves as a small perturbation and creates a significant correction to the eigenstates (but only a small correction to the spectrum). 
This is in contrast to the general case in the Appendix, where the spectrum is nondegenerate, and thus both FSA and $SO(3)$-FSA work well for the projection. 

Finally, we investigate the EE of the Floquet eigenstates in Fig.~\ref{Fig:Fscar_overlap_DTC}(c). 
We find that the eigenstates of $U(T)$ are all highly entangled. 
In contrast, as shown in Fig.~\ref{Fig:Fscar_overlap_DTC}(d), the short-time evolution is dominated by their low-entangled superpositions. 
What is more, the FSA approach works perfectly for the EE, and the $SO(3)$-FSA also provides a good approximation. 
Besides, we have studied a more general case in the Appendix. 
We show that it also possesses the nontrivial Floquet QMBS, and find that the two approximations (FSA and $SO(3)$-FSA) are even better in that case because there is generally no degeneracy unless deliberately designed.

\lettersection{Discussions}
In this work, we study a generalized PXP model whose QMBS states are well described 
by an FSA subspace that is universal for all choices of $\vec{n}$. 
Hence, the FSA subspace is approximately an invariant subspace of the generalized PXP model, which carries an approximate $SO(3)$ symmetry. 
Hence, the evolution in the FSA subspace of a time-dependent system can be solved by a pair of time-dependent rotation axis and angles. 
Utilizing this property, we have much freedom to design the system, and particularly, we engineer a Floquet system with period-tripling revivals in the main text. 
Moreover, the group structure also reveals that the Floquet Hamiltonian is actually captured by another generalized PXP model, which carries nontrivial Floquet QMBS states. 

Despite the resemblance to the prethermal DTC in the literature~\cite{Abanin2017,Else2017,Maskara2021_PRL,Biao2022}, there are important differences between our protocol and the earlier work. 
In particular, the subharmonic response here arises from the continuous symmetry $SO(3)$, while prethermal DTCs spontaneously break a discrete symmetry. 
This work also shows the probability of QMBS states in the nearly-degenerate systems in which QMBS states are not exact eigenstates but their superpositions.

\lettersection{Acknowledgement}
This work is supported by the Research Grants Council of Hong Kong (Grants~No.~CityU~21304720, CityU~11300421, and C7012-21G), and City University of Hong Kong (Project~No.~9610428).  
K.H. is also supported by the Hong Kong PhD Fellowship Scheme. 

\lettersection{Note added}
While finishing up this work, we became aware of an independent work exploring a similar idea in a different setup, which will appear in the same arXiv posting~\cite{Zhicheng}. 

\bibliographystyle{apsrev4-2}
\bibliography{FloquetScar_v5b.bib}

\appendix 
\onecolumngrid

\section*{Supplemental Materials for ``Designing a pre-thermal discrete time crystal with arbitrary periods''}

\begin{figure}[b]
\includegraphics[width=0.6\columnwidth]{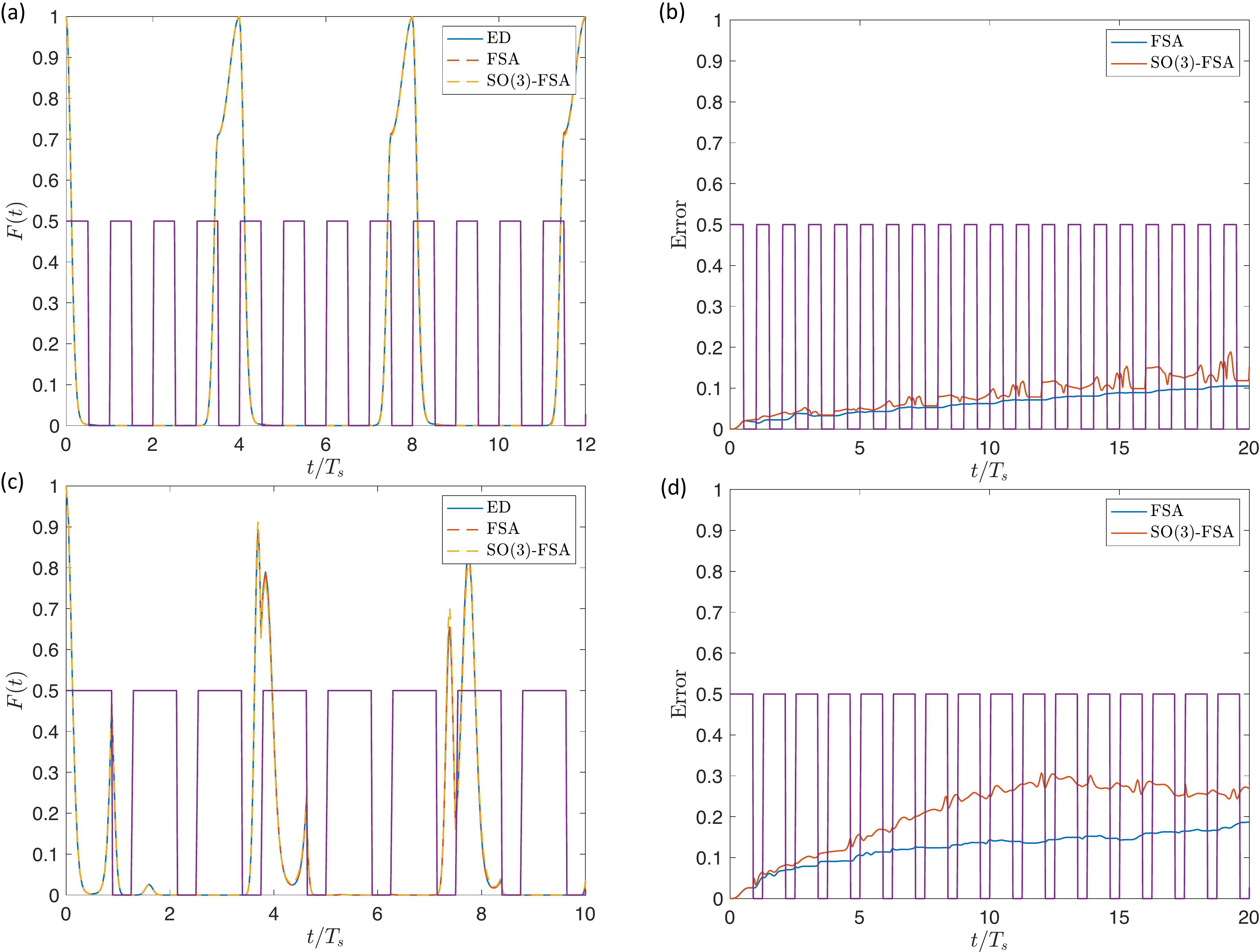}
\caption{\label{Fig:Dynamics_new}
(a) and (b) show the fidelity and the accuracy respectively in an $L=28$ chain ($T_{s}=9.8987$), with parameters $\vec n_1=(\pi/5,0)$, $\vec n_2=(\pi/5+3\pi/4,0)$, $T_1=T_s/2$, and $T_2=T_s/2$. 
(c) and (d) show the fidelity and the accuracy respectively in $L=28$ chain ($T_{s}=9.8987$), with parameters $\vec n_1=(\pi/5,0)$, $\vec n_2=(2\pi/5,\pi/2)$, $T_1=7T_s/8$, $T_2=3T_s/8$. 
The purple line in (a) and (c) equals $1/2$ when $\vec n_1\cdot\vec H$ is turned on and $0$ otherwise.  
}
\end{figure}

\section{Engineering generic Floquet evolutions}
In this section, we study a more general case for the Floquet Hamiltonian in Eq.~\eqref{Eq:Floquet_Hamiltonian}. 
We first discuss a period quadrupling case, which is shown in Fig.~\ref{Fig:Dynamics_new}(a) and \ref{Fig:Dynamics_new}(b) with parameters $\vec n_1=(\pi/5,0)$, $\vec n_2=(\pi/5+3\pi/4,0)$, $T_1=T_s/2$, and $T_2=T_s/2$. Similar to the period-tripling case, the dynamics can be accurately captured by the $SO(3)$ structure in the FSA subspace. 

A more generic scenario of the $SO(3)$ structure can also be found in systems without subharmonic responses. 
Particularly, we take $\vec n_1=(\pi/5,0)$, $n_2=(2\pi/5,\pi/2)$, and $T_1=7T_s/8$, $T_2=3T_s/8$. 
Note that the properties of $\vec n_{1,2}\cdot \vec H$ have been studies in Fig.~\ref{Fig:OverlapEE}. 
We first study the dynamics of $\ket{\mathbb{Z}_2}$. 
In Fig.~\ref{Fig:Dynamics_new}(c), two approximations fit the ED results excellently, and Fig.~\ref{Fig:Dynamics_new}(d) even suggests that the $SO(3)$-FSA approach is better here than the case in the main text. 
Another intriguing phenomenon here is that as the system is not carefully designed, there is never perfect revival, i.e. the corresponding 3D rotation $R(t)$ never becomes the identity.

\begin{figure}[!]
\includegraphics[width=0.6\columnwidth]{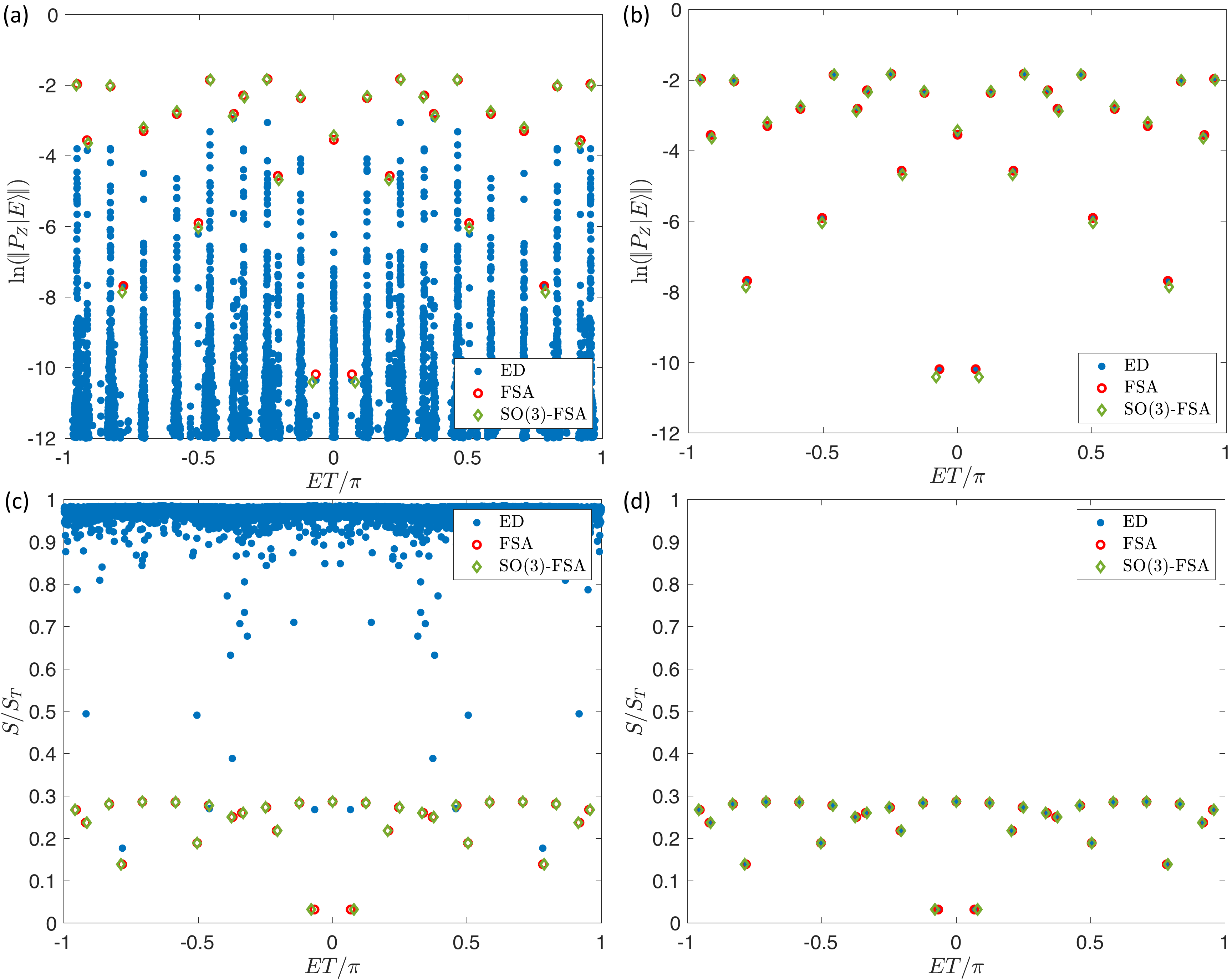}
\caption{\label{Fig:Fscar_overlap} The projection and the EE of the eigenstates derived from three different approaches. 
The ED results in the left and the right panel represent the eigenstates of $F$ and $P_FF P_F$, respectively. 
The parameters used in this figure are identical to those in Fig.~\ref{Fig:Dynamics_new}.}
\end{figure}

In addition, we also compute the Floquet eigenstates of this system, which are shown in Fig.~\ref{Fig:Fscar_overlap}. 
This generic system manifests similar features to the case in the main text, i.e. highly entangled eigenstates form tower structures whose superposition serves as the scar states. However, the most significant difference here is that the $SO(3)$-FSA approach is almost as accurate as the FSA approach, indicating the universal applicability of the $SO(3)$-FSA approach.

\section{The fate of imperfect revivals}
In the previous section, we demonstrate that the revival is generally imperfect without deliberate design. Hence, a natural question is whether these imperfect revivals survive in the thermodynamic limit. 
From the $SO(3)$ point of view, we can ask whether we have
\begin{align}
    \lim_{j\to\infty}\abs{\mel{j,j,\vec z}{U_j(R)}{j,j,\vec z}}=0, 
\end{align}
where $\ket{l,m,\vec n}$ is the eigenstate of $\vec n\cdot J$ with $\vec J^2\ket{l,m,\vec n}=l(l+1)\ket{l,m,\vec n}$ and $\vec n\cdot J\ket{l,m,\vec n}=m\ket{l,m,\vec n}$. Here, $U_l(R)$ is the corresponding $(2l+1)$-dimensional irreducible representation of $R\in SO(3)$.

To start, we first prove the following lemma:
for all $l,m$, and two unit vectors $\vec n,\vec n'$, the following inequality holds
\begin{align}
    \abs{\braket{l,m,\vec n}{l,m,\vec n'}}\leq \left[1+\frac{2m^2\tan^2(\theta/2)}{l(l+1)-m^2}\right]^{-1/2}, 
\end{align}
where $\theta$ is the angle between $\vec n$ and $\vec n'$.
To show this, first note that the inequality is trivial, and the equality holds if $\vec n=\pm\vec n'$. Hence, we only consider that $\vec n$ and $\vec n'$ are not parallel, or equivalently $\sin\theta\neq0$.
Without loss of generality, we set $\vec n=\vec z$. Further, because
\begin{align}
    \abs{\braket{l,m,\vec n}{,m,\vec n'}}
    =\abs{e^{im\phi}\braket{l,m,\vec n}{,m,\vec n'}}
    =\abs{\mel{l,m,\vec n}{e^{-i\phi J_l^z}}{l,m,\vec n'}}
    =\abs{\braket{l,m,\vec n}{l,m,R_z(\phi)\vec n'}},
\end{align}
we set $\vec n'$ in the xoz plane, that is $\vec n'=\cos\theta \vec z+\sin\theta\vec x$.

Now let $\ket{\psi}=\sum_{k}\psi_k\ket{l,k}$ be the eigenvalue of $\vec n'\cdot\vec J_l$ with eigenvalue $m$, and then we have
\begin{align}
    \abs{\braket{l,m,\vec n}{l,m,\vec n'}}=\dfrac{\abs{\psi_m}}{\norm{\ket{\psi}}}. 
\end{align}
We suppose that $\psi_m\neq0$, otherwise the lemma is proved. 
Therefore, we can set $\psi_m=1$, and
\begin{align}
    m= \mel{l,m}{\vec n'\cdot\vec J_l}{\psi} 
    =m\cos\theta+\frac{\sin\theta}2\psi_{m+1}\sqrt{l(l+1)-m(m-1)} 
    +\frac{\sin\theta}2\psi_{m-1}\sqrt{l(l+1)-m(m+1)}, 
\end{align}
which can be rewritten as
\begin{align}
    2m\tan(\theta/2)=&\psi_{m+1}\sqrt{l(l+1)-m(m+1)} 
    +\psi_{m-1}\sqrt{l(l+1)-m(m-1)}.
\end{align}
From the Cauchy inequality, we have
\begin{align}
    \tan^2(\theta/2)\leq\frac{l(l+1)-m^2}{2m^2}\left(\abs{\psi_{m+1}}^2+\abs{\psi_{m-1}}^2\right),
\end{align}
so we can further derive
\begin{align}
    \norm{\ket{\psi}}^2 \geq\left(\abs{\psi_{m+1}}^2+\abs{\psi_{m}}^2+\abs{\psi_{m-1}}^2\right) \geq 1+\frac{2m^2\tan^2(\theta/2)}{l(l+1)-m^2}.
\end{align}
Hence we have
\begin{align}
    \abs{\braket{l,m,\vec n}{l,m,\vec n'}} =\abs{\psi_m}/\norm{\ket{\psi}} 
    \leq\left[1+\frac{2m^2\tan^2(\theta/2)}{l(l+1)-m^2}\right]^{-1/2},
\end{align}
which completes the proof of the lemma.

According to the lemma, it is readily to derive the following theorem:
for an arbitrary nonnegative integer $k$, a rotation $R$ and a unit vector $\vec n$, we have
\begin{align}
    \lim_{l\to\infty}\abs{\mel{l,l-k,\vec n}{U_l(R)}{l,l-k,\vec n}}=\delta_{\vec n,R\vec n}, 
\end{align}
where $\delta_{\vec n,\vec n}=1$ and $\delta_{\vec n,\vec n'}=0$ if $\vec n\neq\vec n'$.
First, noticing that $\abs{\mel{l,l-k,\vec n}{U_l(R)}{l,l-k,\vec n}}=\abs{\braket{l,l-k,\vec n}{l,l-k,R\vec n}}$, according to the lemma, we have
\begin{align*}
    \abs{\mel{l,l-k,\vec n}{U_l(R)}{l,l-k,\vec n}}\leq\left[1+\frac{2m^2\tan^2(\theta/2)}{l(l+1)-m^2}\right]^{-\frac{1}{2}}.
\end{align*}

If $\vec n=R\vec n$, then $\abs{\mel{l,l-k,\vec n}{U_l(R)}{l,l-k,\vec n}}=1$ and the limit equals $1$. If $\vec n\neq R\vec n$, we know that 
\begin{align}
    \lim_{l\to\infty}\left[1+\frac{2(l-k)^2\tan^2(\theta/2)}{l(l+1)-(l-k)^2}\right]^{-1/2}=0,
\end{align}
and therefore we have 
\begin{align}
    \lim_{l\to\infty}\abs{\mel{l,l-k,\vec n}{U_l(R)}{l,l-k,\vec n}}=0.
\end{align}
In conclusion, all imperfect revival vanishes in the thermodynamic limit, whereas the perfect revival is always perfect.

\end{document}